\documentclass[aps,prd,nofootinbib,twocolumn,superscriptaddress]{revtex4}

\usepackage{graphicx}
\usepackage{amsmath}
\usepackage{slashed}
\usepackage{color}

\newcommand{\ie}{{\it i.e., }}

\newcommand{\be}{\begin{equation}}
\newcommand{\ee}{\end{equation}}
\newcommand{\br}{\begin{eqnarray}}
\newcommand{\bea}{\begin{eqnarray}}
\newcommand{\eea}{\end{eqnarray}}
\newcommand{\er}{\end{eqnarray}}
\newcommand{\ba}{\begin{array}}
\newcommand{\ea}{\end{array}}
\newcommand{\bi}{\begin{itemize}}
\newcommand{\ei}{\end{itemize}}
\newcommand{\bn}{\begin{enumerate}}
\newcommand{\en}{\end{enumerate}}
\newcommand{\bc}{\begin{center}}
\newcommand{\ec}{\end{center}}

\newcommand{\beq}{\begin{equation}}
\newcommand{\eeq}{\end{equation}}

\newcommand{\gsim}{\lower.7ex\hbox{$\;\stackrel{\textstyle>}{\sim}\;$}}
\newcommand{\lsim}{\lower.7ex\hbox{$\;\stackrel{\textstyle<}{\sim}\;$}}

\def\mysection#1{{\bf #1.} }

\begin{document}

\title{Galactic Centre GeV Photons from Dark Technicolor}

\author{Matti Heikinheimo}
\affiliation{National Institute of Chemical Physics and Biophysics, R\"avala 10, 10143 Tallinn, Estonia}

\author{Christian Spethmann}
\affiliation{National Institute of Chemical Physics and Biophysics, R\"avala 10, 10143 Tallinn, Estonia}

\date{\today}

\begin{abstract}
We present a classically scale-invariant model with a confining dark sector, which is coupled to the Standard Model through the Higgs portal. The galactic centre gamma ray excess can be explained in this model by collision-induced dark matter decays to b-quarks. We discuss the possibility to obtain the dark matter relic density through thermal freeze-out, which however requires excessive fine-tuning. We then instead focus on a freeze-in scenario and perform detailed calculations and a parameter scan. We find that the observed relic density and the gamma ray excess can be explained by a wide range of parameters in this model.
\end{abstract}

\maketitle

\section{Introduction}

During the last decades, much effort has been spent to search for evidence of dark matter annihilation in the Milky Way. Gamma rays offer the most intriguing possible signal, because they travel unperturbed through the interstellar medium from the source to the detectors, and can be unambiguously observed. For some years now it has been know that there is an apparent excess of 1-3 GeV gamma rays in Fermi-LAT data from the galactic centre \cite{Goodenough:2009gk, Hooper:2010mq, Boyarsky:2010dr, Hooper:2011ti, Abazajian:2012pn, Gordon:2013vta, Abazajian:2014fta} and the inner regions of the galaxy \cite{Hooper:2013rwa, Huang:2013pda}. 

In addition, it has been noted that data from the PAMELA experiment \cite{Adriani:2008zq, Adriani:2010rc, Adriani:2012paa} exhibit a 40\% excess of antiprotons in the 1-3 GeV energy range over expectations from cosmic ray propagation models \cite{Evoli:2011id, DiBernardo:2009ku, Jin:2014ica}. Miraculously, both excesses can be generated by the same \mbox{30-40 GeV} dark matter annihilating into $b\bar{b}$ with a thermally averaged cross section of \mbox{$\langle\sigma v\rangle \approx (1.4-2) \, 10^{-26} \ {\rm cm}^3/{\rm s}$}~\cite{Daylan:2014rsa, Hooper:2014ysa}. A slightly heavier dark matter mass was preferred in~\cite{Calore:2014xka}. As noted in~\cite{Basak:2014sza,Modak:2013jya,Alves:2014yha,Alvares:2012qv}, this setup naturally occurs in Higgs-portal dark matter models in the presence of a scalar resonance. 

In this paper we show how the observed photon and antiproton excesses can be explained by the collision-induced decays of composite dark matter in a classically scale-invariant model. In this setup~\cite{Hur:2007uz,Hur:2011sv,Heikinheimo:2013fta,Holthausen:2013ota}, here referred to as \emph{Dark Technicolor}, the electroweak scale is generated dynamically in a strongly interacting hidden sector and mediated to the SM via a Higgs portal interaction with a messenger singlet. The stable dark matter candidate is a composite state, {\it i.e.} a dark technipion or dark technibaryon.

The coupling between the visible sector and the dark sector through the Higgs portal is necessarily weak to avoid LHC and direct detection constraints. This means that thermal freeze-out is not a viable way to generate a sufficiently small relic density unless the dark matter annihilation process is enhanced by an $s$-channel resonance, which requires a fine-tuned mass spectrum. Instead, we propose that the observed dark matter can be generated via freeze-in if the couplings of the messenger scalar are sufficiently small.

We will present our setup in section \ref{sec:model}. In section \ref{sec:gammaray} we discuss in detail the origin of the gamma-ray signal. In section \ref{sec:freezein} we discuss dark matter generation through the freeze-in mechanism in the early universe. In section \ref{sec:results} we discuss the results of our numerical analysis and in section \ref{sec:conclusions} we conclude and discuss further implications of the model.

\section{The Dark Technicolor Model} 
\label{sec:model}

The dark technicolor model consists of a confining dark sector, \ie dark techniquarks $Q$ charged under a non-Abelian gauge group $SU(N)_{\rm TC}$, and a scalar messenger $S$ that is a singlet under all gauge groups. The model, including the SM, is taken to be classically scale invariant, so that any explicit mass terms are absent at tree level. As the dark technicolor sector becomes confining at a scale $\Lambda_{\rm TC}$, the messenger field obtains a vacuum expectation value, which in turn generates an effective negative mass squared term for the Higgs via a Higgs portal coupling. Electroweak symmetry breaking then proceeds as in the SM.

The Lagrangian of the model is given by
\bea
\mathcal{L} &=&|D_\mu H|^2 + |\partial_\mu S|^2-\frac14 F^{a \mu\nu}F^a_{\mu\nu} + \sum_{i=1}^{N_f}\bar{Q_i}i\slashed{D}Q_i \nonumber \\ &&- \lambda_h |H|^4 - \frac14\lambda_S |S|^4 + \lambda_{Sh}|S|^2|H|^2 \nonumber \\ &&- (\sum_{i=1}^{N_f}y_{Q_i}S\bar{Q_i}(1+a_i\gamma^5)Q_i + {\rm h.c.}) + \mathcal{L}_{\rm SM, m_H=0},
\label{DarkTCLagrangian}
\eea
where $\mathcal{L}_{\rm SM, m_H=0}$ contains the SM gauge and fermion sectors, including the usual Yukawa couplings with the Higgs, but no explicit Higgs mass term. $H$ is the Higgs doublet, $H=\frac{1}{\sqrt{2}}(0,h+v)$ in the unitary gauge. $F^a_{\mu\nu}$ is the field strength of the dark technicolor gauge group, and $N_f$ is the number of flavors in the hidden sector. The singlet messenger scalar $S$ has Yukawa couplings with the hidden sector quarks, parametrized by $y_{Q_i}$, which are assumed to be flavour diagonal. We have included a parity-violating term in the Yukawa-couplings, parametrized by the constants $a_i$, which is required for the pion decay through the scalar messenger.
 
The confinement and chiral symmetry breaking results in Pseudo-Nambu-Goldstone bosons---massive dark technipions, which are our dark matter candidate. In the absence of explicit chiral symmetry breaking terms these particles would be massless, but here the chiral symmetry of the dark technicolor sector is explicitly broken by the Yukawa couplings. The pion mass is estimated to be~\cite{GellMann:1968rz}
\be
m_\pi^2\approx m_Q\frac{\Lambda_{\rm TC}^3}{f_\pi^2}=y_Qv_S\frac{\Lambda_{\rm TC}^3}{f_\pi^2},
\label{eq:pionmass}
\ee
where $f_\pi$ is the dark technipion decay constant and $\Lambda_{TC}$ is the dark technicolor confinement scale. $v_S$ is the vacuum expectation value of the messenger field, given by minimising the scalar potential in (\ref{DarkTCLagrangian}) as
\be
v_S=\left(\frac{y_Q\lambda_h}{\lambda_S\lambda_h-\lambda_{Sh}^2}\right)^{\frac13}\Lambda_{\rm TC},
\label{eq:v_S}
\ee
where $\lambda_{Sh}$ is the portal coupling and $\lambda_h\approx 0.13$ is the Higgs self coupling. The vev of the Higgs field is given as
\be
v_{EW}=\sqrt{\frac{\lambda_{Sh}}{\lambda_h}}v_S,
\label{eq:vevs}
\ee
and the mass matrix of the scalar sector is
\be
\mathcal{M}=\begin{pmatrix} 3\lambda_S-\frac{\lambda_{Sh}^2}{\lambda_h} && -2\sqrt{\frac{\lambda_{Sh}^3}{\lambda_h}} \\ -2\sqrt{\frac{\lambda_{Sh}^3}{\lambda_h}} && 2\lambda_{Sh} \end{pmatrix} v_S^2,
\label{eq:massmatrix}
\ee
so that in the zero-mixing limit the Higgs mass returns the SM value 
\be
m_h=\sqrt{2\lambda_h}v_{EW},
\ee
and the mass of the messenger scalar is
\be
m_S=\sqrt{3\lambda_S-\frac{\lambda_{Sh}^2}{\lambda_h}}v_S.
\label{eq:scalarmass}
\ee
The mixing angle between the messenger and the Higgs is obtained by diagonalising the mass matrix (\ref{eq:massmatrix}) as
\be
\tan(2\theta_{Sh})=\frac{4\lambda_{Sh}^{\frac32}}{\sqrt{\lambda_h}\left(\lambda_{Sh}\left(2-\frac{\lambda_{Sh}}{\lambda_h}\right)-3\lambda_S\right)}.
\label{eq:mixing angle}
\ee
We will work in the limit where the mixing angle is small and hence the Higgs potential has the same form as in the SM.

\section{The Gamma-ray Signal}
\label{sec:gammaray}

The dark matter in our model consists of the stable technicolor mesons and baryons, which carry a conserved quantum number and are therefore protected from decaying. This symmetry charge can be either the dark sector baryon number, or dark sector flavor symmetries if the Yukawa couplings to the messenger scalar are strictly diagonal. 

As shown in \cite{Buckley:2012ky}, it is possible to stabilise some of the mesons of the theory through dark baryon number if the fermions transform under the adjoint representation of the dark technicolor gauge group. In the simplest realisation of this scenario, the chiral symmetry breaking structure is $SU(2 N_f) \to Sp(2 N_f)$,
resulting in 3 unstable and 2 stable pseudo-Goldstone bosons for $N_f=2$. 

Alternatively, QCD-like chiral symmetry breaking with strictly diagonal Yukawa couplings results in 2 stable ($\pi_1, \bar{\pi}_1$, corresponding to $\pi^\pm$ in the SM) and one unstable ($\pi_0$) meson. For concreteness, we concentrate our numerical analysis on this scenario, \ie $N_f=2$ flavours in the fundamental representation of the technicolor gauge group with exactly diagonal Yukawa couplings. However,  the general conclusions of the paper are just as valid for other symmetry breaking structures, if one compensates for the different group factors by changing the couplings by corresponding $\mathcal{O}(1)$ factors.

\begin{figure}[t]
\begin{center}
\includegraphics[width=0.2\textwidth]{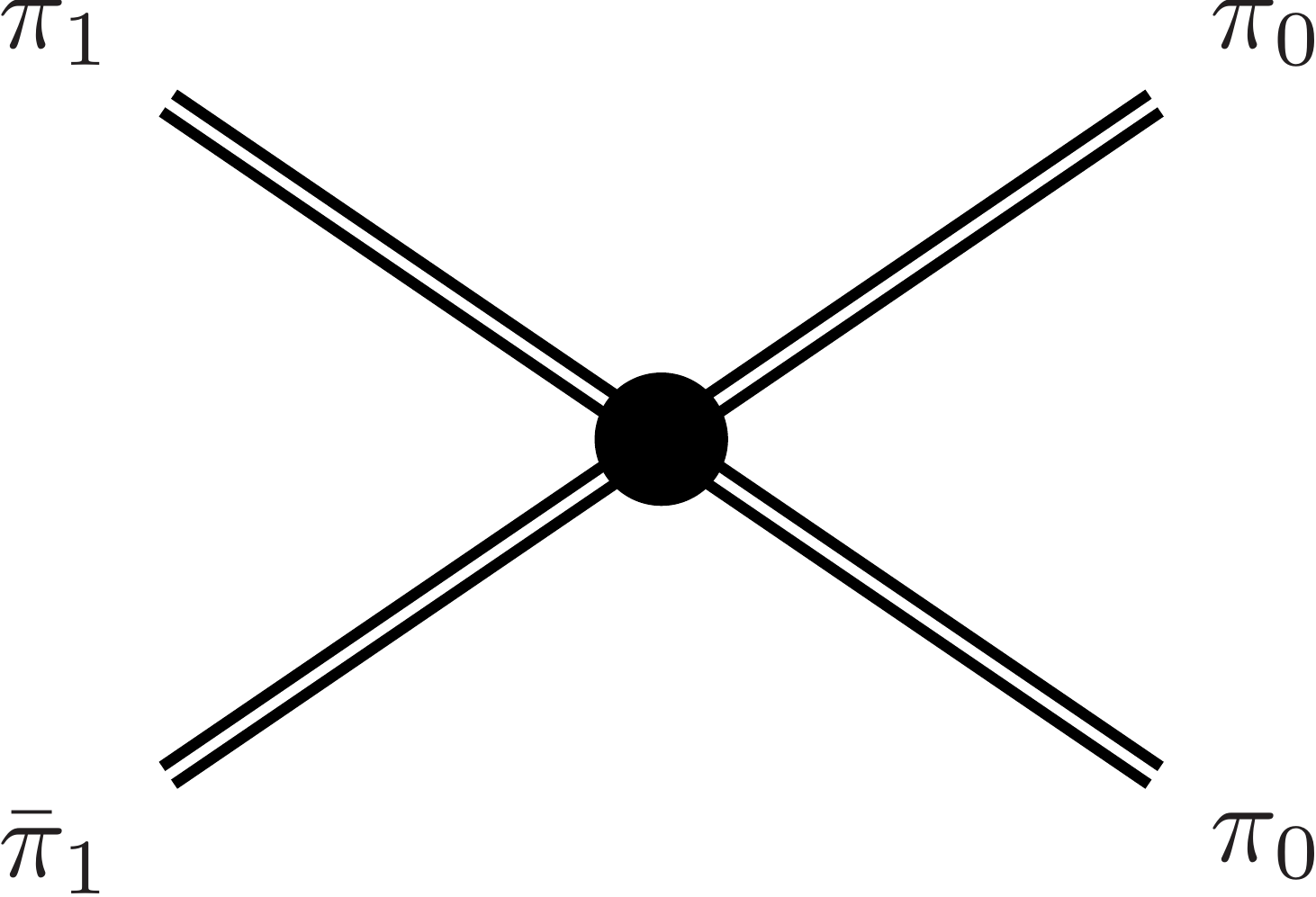}
\caption{Scattering of stable $\pi_1$ NGBs into unstable $\pi_0$ NGBs.}
\label{fig:scatter}
\end{center}
\end{figure}

A dark matter particle with a mass of 31-40 GeV and an annihilation cross section of $(1.4-2)10^{-26} \ {\rm cm}^3/{\rm s}$ into $b\bar{b}$ can fit the gamma ray signal very well \cite{Daylan:2014rsa}. In our model, however, the annihilation proceeds through the intermediate state of unstable technipions, resulting in four $b$-quarks per DM collision, \mbox{$\pi_1\bar\pi_1\to\pi_0\pi_0\to 4b$}. Here $\pi_1\sim|\bar{Q}_2Q_1\rangle$ is the dark matter particle, which is stable due to the unbroken flavour symmetry of the dark technicolor sector and $\pi_0\sim\frac{1}{\sqrt{2}}(|\bar{Q}_1Q_1\rangle-|\bar{Q}_2Q_2\rangle)$ is the unstable pion that decays into $b\bar{b}$ through the Higgs portal. In order to reproduce the spectrum of the resulting $b$-quarks, we set the mass of the technipions to $\sim 80$~GeV. Since dark matter in the galatic centre is non-relativistic (v/c $\sim 10^{-3}$), the expected photon spectrum from the decay of a 80 GeV dark pion to $b\bar{b}$ is identical to the spectrum of direct s-channel annihilation of 40 GeV dark matter to $b \bar{b}$. For a more detailed discussion of the resulting gamma-ray spectrum in a similar scenario, see~\cite{Abdullah:2014lla}. Other models where the gamma-ray signal originates from annihilations in a hidden sector have been previously considered in~\cite{Berlin:2014pya,Hooper:2012cw,Ko:2014gha,Boehm:2014bia,Martin:2014sxa,Cline:2014dwa, Ko:2014loa,Freytsis:2014sua}.

In order to reproduce the observed gamma-ray excess, we will then set the self scattering cross section \mbox{$\sigma(\pi_1\bar\pi_1\to\pi_0\pi_0)$} to the desired range of \mbox{$(2.8-4)10^{-26} \ {\rm cm}^3/{\rm s}$}. This number is a factor of 2 larger than the number given in~\cite{Daylan:2014rsa}, since in our model the dark matter mass is larger by a factor of two, as discussed above. Therefore the number density of our dark matter is reduced by a factor of two, resulting in a reduction of the annihilation rate by a factor of four. But as each annihilation results in four $b$-quarks instead of two, this results in reduction of the final flux of SM particles by a factor of two, which is then compensated by the larger cross section.

The thermally averaged pion self scattering cross section is given as~\cite{Buckley:2012ky}
\be
\langle\sigma v\rangle \approx \frac{9m_\pi^2}{16\pi^\frac32 f_\pi^4\sqrt{x}},
\label{eq:self scattering xsec}
\ee
where $x=m_\pi/T$ is the pion mass divided by the temperature of the DM. Taking $m_\pi=80$ GeV, $x\sim 10^3$ and requiring \mbox{$\langle \sigma v\rangle \sim (2.8-4)\times 10^{-26}\ {\rm cm^3/s}$} we can solve for $f_\pi$, resulting in \mbox{$f_\pi\sim 280-300$ GeV.} 

The pion decay constant is not perturbatively calculable, but since it is found to be in the right ballpark between the pion mass and the confinement scale, it is reasonable to assume that by choosing the right gauge group and fermion representations, a theory can be found which reproduces the correct signal strength. 

\begin{figure}[t]
\begin{center}
\includegraphics[width=0.3\textwidth]{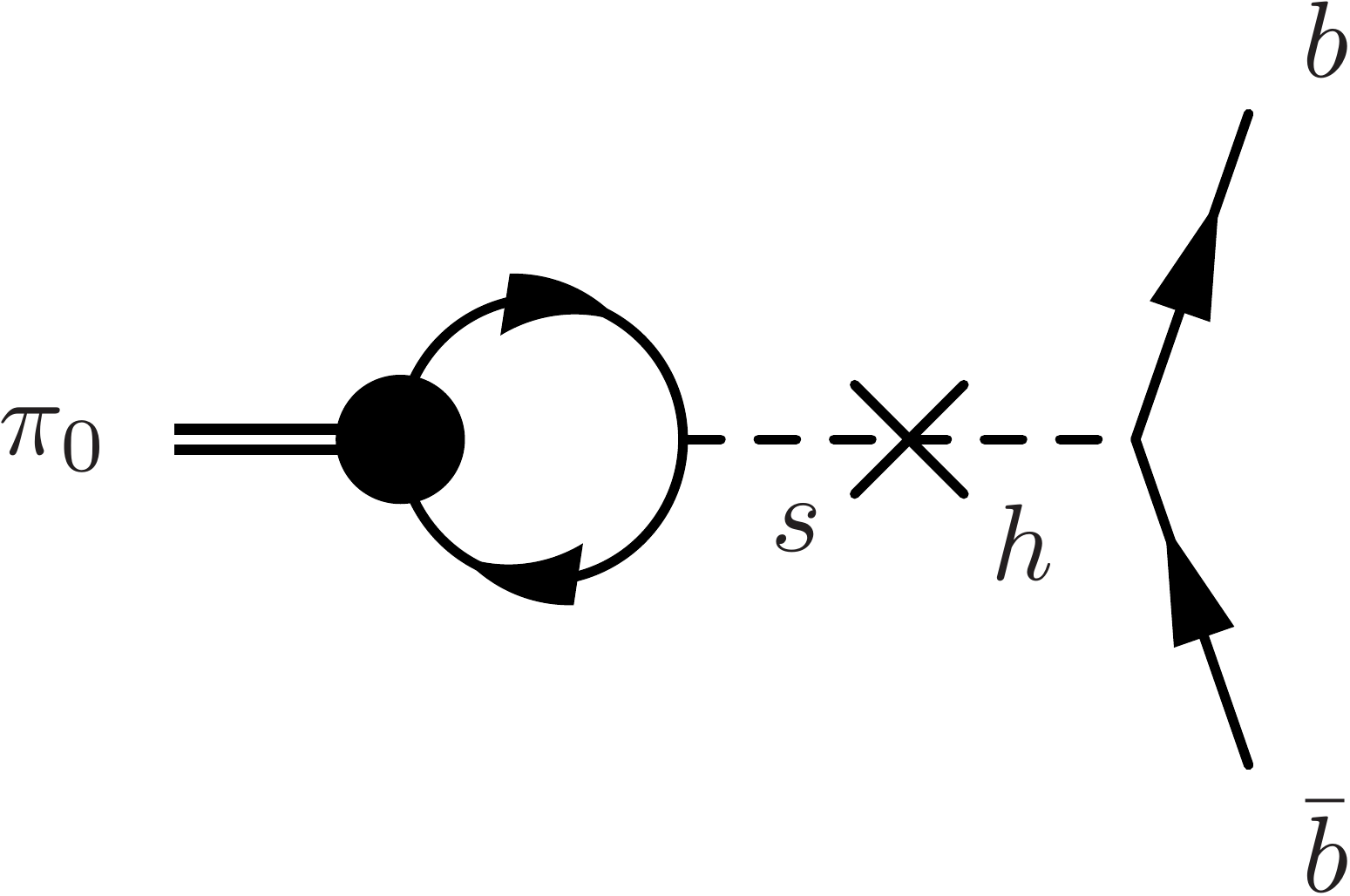}
\caption{Decay of the unstable $\pi_o$ NGB to $b \bar{b}$ through the Higgs portal.}
\label{fig:decay}
\end{center}
\end{figure}

The dominant contribution to the decay width of the unstable pion is the $s$-channel diagram shown in figure \ref{fig:decay}. The resulting decay width is given by
\be
\Gamma(\pi_0\to b\bar{b})=\frac{C_\pi^4a^2y_Q^2\sin^2\theta_{Sh}m_b^2m_\pi}{8\pi v_{EW}^2(m_\pi^2-m_{\min}^2)^2}\left(1-\frac{4m_b^2}{m_\pi^2}\right)^\frac32.
\label{eq:pion width}
\ee
where $m_{\min}=\min(m_H,m_S)$ and $C_\pi$ is an unknown constant with dimension of mass, describing the strength of the $\pi S$ coupling induced by the Yukawa coupling. The decay amplitude is proportional to the parity-violating part of the Yukawa-coupling $ay_Q$, since the pseudoscalar pion has to decay through the scalar mediator, which is forbidden if parity is exactly conserved. We will take $a\approx 1$ for simplicity. This will induce a small parity violation also to the SM sector through the Higgs mixing, but the effects are suppressed by the small portal coupling and are thus phenomenologically negligible.

\section{Freeze-in vs.~freeze-out}
\label{sec:freezein}

We will in the following first discuss the possibility to obtain the correct relic density in a standard freeze-out scenario, and then for the rest of the paper concentrate on thermal freeze-in.

\subsection{Problems with Thermal Freeze-Out}

\begin{figure}[t]
\begin{center}
\includegraphics[width=0.3\textwidth]{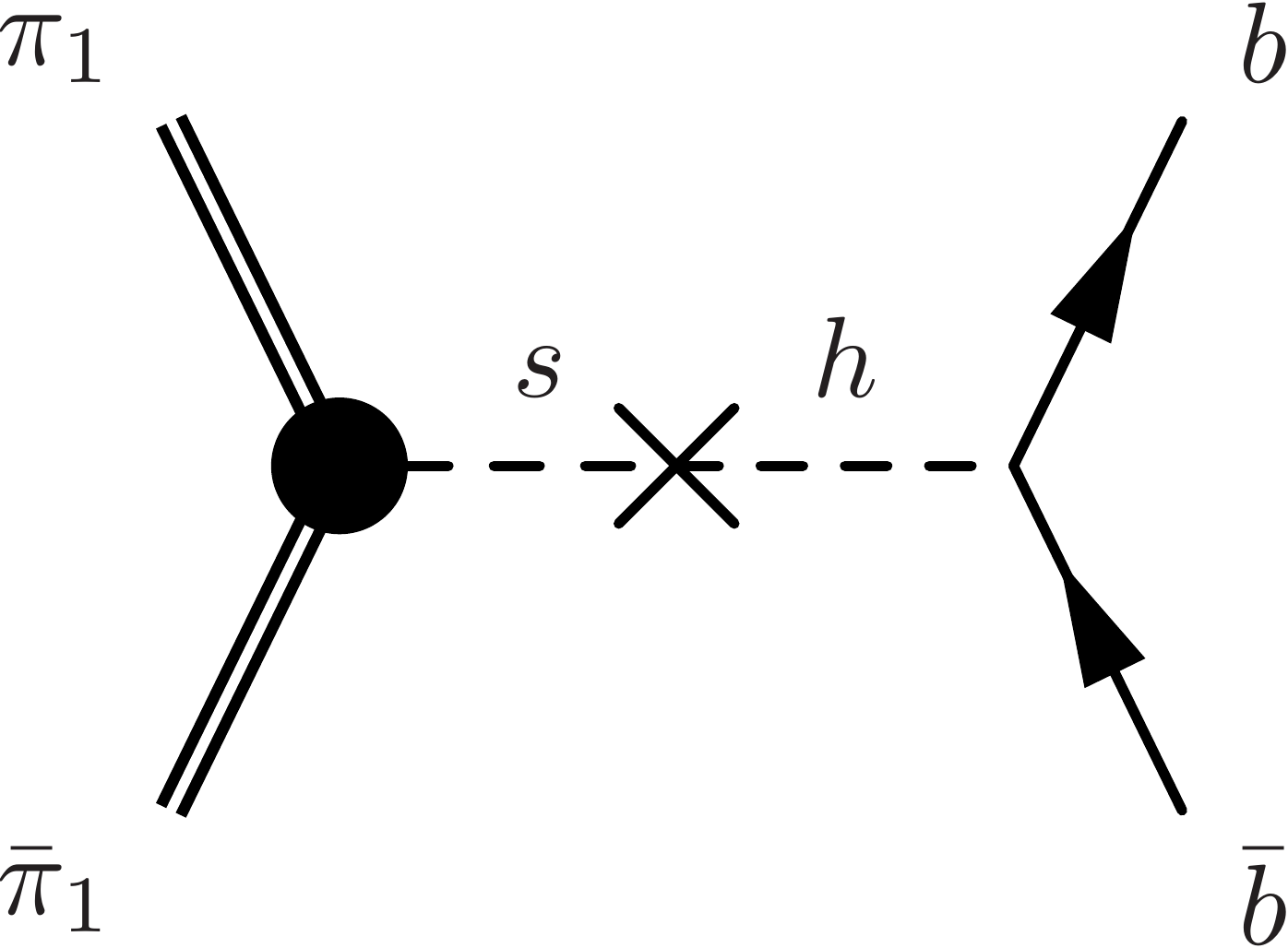}
\caption{Annihilation of the stable $\pi_1$ NGBs to $b \bar{b}$ through the Higgs portal in a thermal freeze-out scenario.}
\label{fig:portal}
\end{center}
\end{figure}

In addition to the pion self scattering and the following decay of the unstable pions, the pions can directly annihilate into $b\bar{b}$ through the $s$-channel diagram shown in figure \ref{fig:portal}. If the lifetime of the unstable pions is large compared to the timescale of the thermal freeze-out process, the \mbox{$2\to2$} self scattering does not change the particle number in the hidden sector and therefore can not be used to set the thermal relic abundance in the freeze-out scenario. Then the relic abundance is set by the thermally averaged cross section of the $s$-channel annihilation process, given by

\be
\langle \sigma v \rangle = \frac{C_\pi^2y_Q^2\sin^2\theta_{Sh}m_b^2}{4\pi v_{EW}^2(4m_\pi^2-m_{\min}^2)^2}\left(1-\frac{m_b^2}{m_\pi^2}\right)^\frac32.
\ee
This is also the case if this cross section is dominant over the pion self scattering process. Here $C_\pi$ is a constant with dimension of mass, describing the strength of the $\pi \pi S$ coupling. In our analysis we have taken \mbox{$C_\pi=f_\pi$}, since the magnitude of this coupling is set by the strong dynamics of the dark technicolor sector. In this case the annihilation process is \mbox{$\pi_1\bar{\pi}_1\to b\bar{b}$} instead of the \mbox{$\pi_1\bar{\pi}_1\to\pi_0\pi_0\to bb\bar{b}\bar{b}$,} so we set the pion mass to \mbox{$m_\pi=40$ GeV} in this section.

\begin{figure}[t]
\begin{center}
\includegraphics[width=0.5\textwidth]{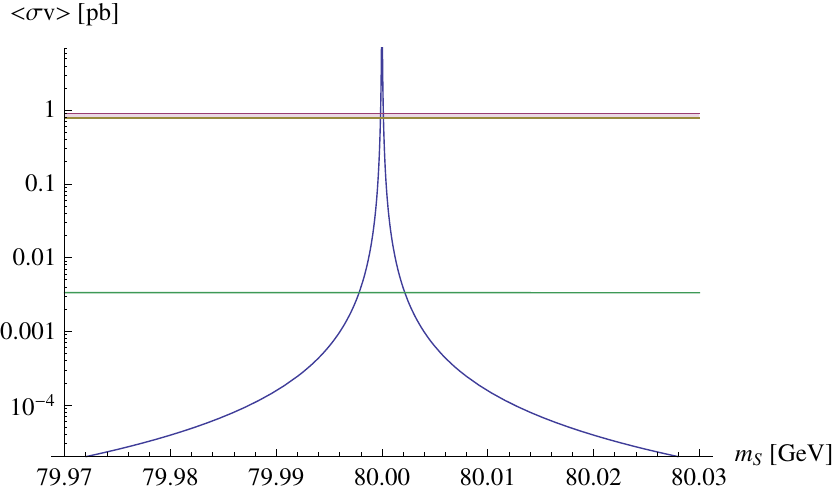}
\caption{Cross-section of the portal interaction from figure \ref{fig:portal} as a function of the messenger scalar mass for $\lambda_{Sh}=5\times10^{-4}$ and \mbox{$C_\pi=f_\pi=\Lambda_{\rm TC}/(4\pi)$}. The horizontal shaded region shows the required value to reproduce the observed relic abundance, and the horizontal green line is the pion self scattering cross section.}
\label{fig:resonance}
\end{center}
\end{figure}

The strength of the portal coupling is restricted by collider constraints from Higgs boson observables and through the non-observation of dark matter at direct detection experiments. If the messenger scalar is lighter than the Higgs, the most stringent limit is set by LEP~\cite{Barate:2003sz,Abbiendi:2002qp}, implying $\sin\theta_{Sh}<0.1$. Therefore the annihilation cross section is usually small, so that the relic abundance in the freeze-out scenario is too large. However, if \mbox{$m_s \approx 2 m_\pi$,} there is a resonant enhancement of the scattering cross section. In this case it is possible to enhance the annihilation cross section enough, so that it will produce the correct relic abundance, but this requires fine-tuning between the pion mass and the mass of the messenger scalar, to the precision of \mbox{$\mathcal{O}(10^{-4})$ GeV} as shown in figure \ref{fig:resonance}. If one is willing to accept this level of fine-tuning, the resonant annihilation process can then be used to create the galactic center gamma ray excess. However, there is no \emph{a priori} reason to expect such fine-tuned mass spectrum, and we will not consider it further in this paper.

In the absence of the resonance, the annihilation process through the Higgs portal is weak and will freeze out early, producing an overabundance of dark matter. In this case, however, the self scattering of the pions dominates over the annihilation. The horizontal green line in figure \ref{fig:resonance} shows the self scattering cross section for the choice of parameters \mbox{$f_\pi=\Lambda_{\rm TC}/(4\pi), \lambda_{Sh}=5\times10^{-4}$.} As is seen in the figure, the pion scattering cross section is stronger than the annihilation process unless $m_S$ is very close to the resonance. Therefore, after freeze-out of the annihilation process, the pions will continue to scatter into each other, keeping the stable and unstable pions in thermal equilibrium. If the lifetime of the unstable species is short compared to the freeze-out of the self scattering process, the self scattering will effectively act as annihilation and could perhaps be used to produce the observed relic abundance. However, the thermally averaged pion self scattering cross section is temperature dependent, as shown in equation (\ref{eq:self scattering xsec}). Thus, if we select the cross section at the freeze-out temperature so that it can produce the observed relic abundance, the much lower temperature in the galactic centre today will result in a too weak gamma ray signal compared to the observed flux. We will therefore not work out the details of this scenario here.

In~\cite{Hochberg:2014dra} it was proposed that the thermal freeze-out abundance of a strongly interacting dark matter sector could be set by the $3\to2$ scattering processes in the hidden sector. While this scenario could possibly be realized in our model, it likely requires additional couplings between the hidden and visible sectors to facilitate efficient heat transfer. Thus we will not consider this possibility further in this paper.

\subsection{Freeze-In Scenario}

We now instead assume that the portal coupling is weak, and that no dark matter is produced through reheating after inflation. The DM abundance is then generated through the freeze-in mechanism, via the scattering processes induced by the Higgs portal, $VV\to\pi\pi$ and $hh\to\pi\pi$, where $V$ refers to the $Z$ and $W$-bosons of the SM, and $h$ is the Higgs boson. The dark technicolor sector then consists of two species of pions, the stable $\pi_1$ that is protected by the unbroken flavor symmetry, and the unstable $\pi_0$ that can decay into SM via the Higgs portal interaction. Both species are created in equal numbers in the scattering processes described above. The Higgs portal coupling has to be very weak in order to facilitate the freeze-in scenario, and thus the lifetime of the unstable pion is quite long, of the order of $\mathcal{O}(10^{-2})$ s, but still short enough not to interfere with nucleosynthesis. After this time, the unstable pions decay into SM via $\pi_0\to b\bar{b}$, and the dark sector from there on consists only of the stable species $\pi_1$. In the galaxy center, where the local DM density is larger, the DM particles will self scatter into the unstable species, which then decays into $b\bar{b}$, resulting in the observed gamma-ray excess. 

The freeze-in scenario in the context of the Higgs portal coupling has been discussed in~\cite{Chu:2011be}. Our model is slightly more complicated, but following their discussion we can reproduce the relevant parts of the analysis. As the dark matter mass in our model is above $\frac12 m_H$, the relevant processes for producing the DM abundance are the scattering processes $VV\to\pi\pi$ and $hh\to\pi\pi$. In our model the relevant cross sections are given as
\bea
\sigma(hh\to\pi\pi) &=& \frac{9m_H^4\sin^2\theta_{Sh}y_Q^2C_\pi^2}{8\pi v_{EW}^2s(s-m_{\min}^2)^2}\sqrt{\frac{s-4m_\pi^2}{s-4m_H^2}},\nonumber \\
\sigma(VV\to\pi\pi) &=& \frac{\sin^2\theta_{Sh}y_Q^2C_\pi^2}{72\pi v_{EW}^2s(s-m_{\min}^2)^2}\sqrt{\frac{s-4m_\pi^2}{s-4m_V^2}}\nonumber \\
&~&\times(s^2-4sm_V^2+12m_V^4),
\label{freeze in cross sections}
\eea
where $V=Z,W$. Again we have taken $C_\pi\approx f_\pi$, and made the approximation that the scattering amplitude is dominated by the lighter of the mass eigenstates $S$ or $h$ in the $s$-channel, neglecting the interference between the two diagrams. This is a good approximation in the limit of small mixing and large mass splitting between the eigenstates, which is assumed here. For a discussion of the phenomenology in the case of approximate mass degeneracy and nearly maximal mixing see~\cite{Heikinheimo:2013cua}.

As discussed in~\cite{Chu:2011be}, the number density of the dark matter is then given by
\be
Y=\left. c\frac{\gamma}{sH}\right|_{T=m_\pi},
\ee
where c is a constant of order of unity, $s$ is the entropy density, $H$ is the Hubble constant and $\gamma$ for each process $ab\to cd$ is defined as
\be
\gamma(ab\to cd) = \frac{T}{64\pi^4}\int_{s_{\min}}^\infty ds\sqrt{s}\hat{\sigma}(s)K_1\left(\frac{\sqrt{s}}{T}\right),
\label{eq:gamma connect}
\ee 
with $s_{\min}= \max((m_a+m_b)^2,(m_c+m_d)^2)$. Here $K_1$ is the modified Bessel function of the second kind and the reduced cross section $\hat{\sigma}$ is given by
\be
\hat{\sigma}(ab\to cd)=\frac{g_a g_b}{c_{ab}}\frac{2((s-m_a^2-m_b^2)^2-4m_a^2m_b^2)}{s}\sigma,
\ee
where $g_{a,b}$ is the number of degrees of freedom of the particle species $a$ or $b$, $c_{ab}=2$ if $a$ and $b$ are identical, and otherwise $c_{ab}=1$, and $\sigma$ is the cross section given in (\ref{freeze in cross sections}). For further details we refer the reader to~\cite{Chu:2011be}.

\section{Numerical Results and Analytic Approximations}
\label{sec:results}

\begin{figure*}[t]
\begin{center}
\includegraphics[width=0.49\textwidth]{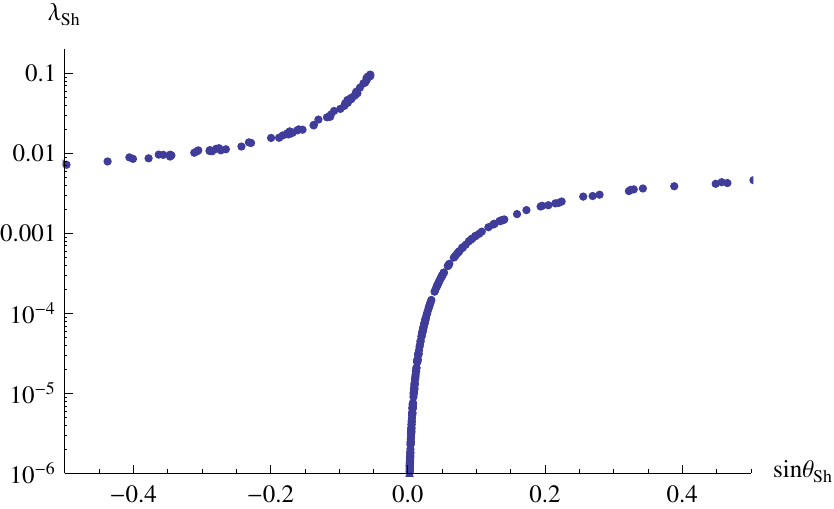}
\includegraphics[width=0.49\textwidth]{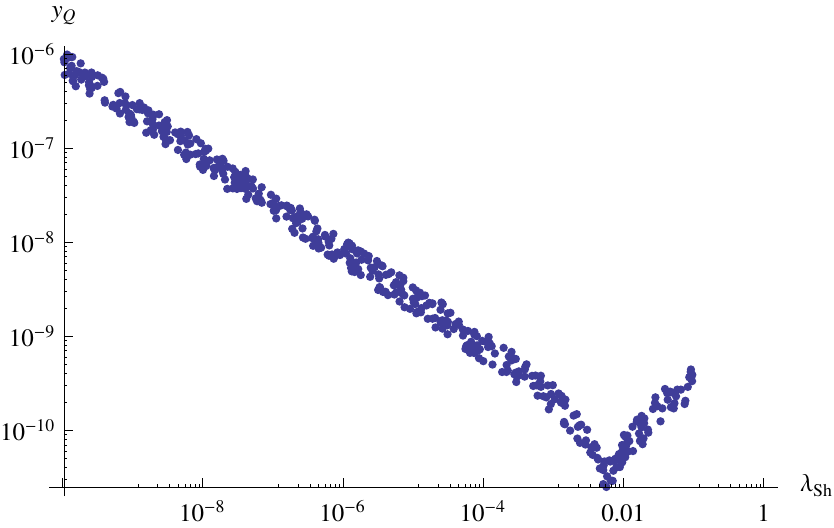}
\includegraphics[width=0.49\textwidth]{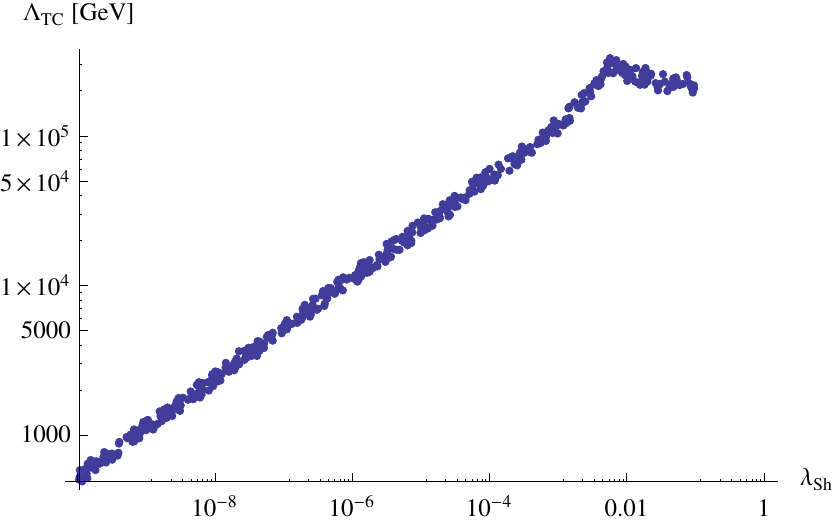}
\includegraphics[width=0.49\textwidth]{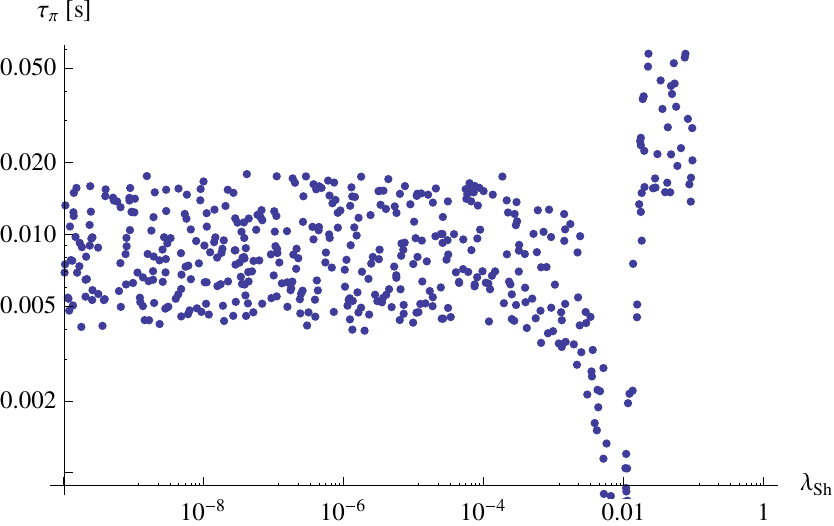}
\caption{Upper left: The portal coupling $\lambda_{Sh}$ as a function of the mixing angle $\sin\theta_{Sh}$.
Upper right: The dark sector Yukawa coupling $y_Q$ as a function of the portal coupling $\lambda_{Sh}$.
Lower left: The dark technicolor confinement scale $\Lambda_{\rm TC}$ (in units of GeV) as a function of the portal coupling $\lambda_{Sh}$.
Lower right: The lifetime of the unstable pion $\tau_{\pi_0}$ (in units of second) as a function of the portal coupling $\lambda_{Sh}$.}
\label{scatterplots}
\end{center}
\end{figure*}

Using the results obtained above, we perform a numerical scan of the parameter space of the model. We use equations (\ref{eq:pionmass})-(\ref{eq:mixing angle}) to solve for the remaining parameters in terms of $f_\pi, m_\pi, \lambda_{Sh}$ and $y_Q$. We fix \mbox{$m_\pi=80$ GeV} and use equation (\ref{eq:self scattering xsec}) to fix $f_\pi$ as explained above. We then vary the parameters $\lambda_{Sh}$ and $y_Q$ in the range \mbox{$[10^{-12},10^{-1}]$} and $f_\pi$ in the range \mbox{$(278-304)$ GeV,} given by equation (\ref{eq:self scattering xsec}). The rest of the parameters in the model \mbox{$(\lambda_S,\Lambda_{\rm TC},m_S,v_S,\theta_{Sh})$} are then given by equations (\ref{eq:pionmass})-(\ref{eq:mixing angle}), and are treated in the numerical analysis as functions of the input variables $f_\pi, m_\pi, \lambda_{Sh}$ and $y_Q$. For each parameter space point we perform the integral in equation (\ref{eq:gamma connect}) numerically to obtain the dark matter relic density. We keep the parameter space points where the relic density is within $\frac12$ to 2 times the observed value, to get an estimate of the typical values for the model. The results are shown in figure \ref{scatterplots}.

The first observation from the upper left plot in figure \ref{scatterplots} is that the mixing angle changes sign at around \mbox{$\lambda_{Sh}\sim 0.005$.} This corresponds to change in the ordering of the masses of the messenger scalar and the Higgs. When $\lambda_{Sh}$ is below the threshold value, the messenger scalar is always lighter than the Higgs, and vice versa. When we approach the threshold, the scalars become degenerate in mass and the mixing angle becomes large. Throughout our analysis we have assumed that cross sections such as equation (\ref{freeze in cross sections}) are dominated by the exchange of the lighter of the scalar mass eigenstates, and that any interference effects can be neglected. In the presence of degenerate masses and large mixing angle this assumption obviously does not hold. Thus, the results in the region $\lambda_{Sh}\sim[0.001,0.03]$ may not be reliable. Furthermore, this region is phenomenologically problematic, since large mixing between the messenger scalar and the Higgs will affect the couplings of the Higgs and would also result in direct production of the messenger scalar in the LHC. This region of parameter space should thus be considered unphysical, or ruled out by experiment.

In the limit of small mixing, there are some features that can be understood simply from the analytical results derived above. In the following, we will discuss the approximations that hold in this limit. However, we have carried out the numerical analysis leading to the results in figure \ref{scatterplots} without relying on any of the approximations discussed below.

First, from the scattering cross sections of equation (\ref{freeze in cross sections}) we see that the relic density is mainly determined by the product $\sin^2\theta_{Sh}y_Q^2$, since $C_\pi\approx f_\pi$ is effectively fixed by equation (\ref{eq:self scattering xsec}), and the pion mass is fixed from the observation of the galactic center gamma ray spectrum. Thus, in order to produce the desired relic density, this product has to be nearly constant, \mbox{$\sin^2\theta_{Sh}y_Q^2\sim6\times10^{-22}$.} From equation (\ref{eq:mixing angle}) we can then solve, in the limit $\theta_{Sh}\ll1$
\be
\sin\theta_{Sh}\approx\frac{2\lambda_{Sh}^{\frac32}}{\sqrt{\lambda_h}(2\lambda_{Sh}-3\lambda_S)}\approx\sqrt{\frac{\lambda_{Sh}}{\lambda_h}},
\ee
where in the last step we have used $\lambda_S\ll\lambda_{Sh}$, which is a good approximation in the small $\lambda_{Sh}$ limit as we will see below. Now, since $\sin\theta_{Sh}y_Q$ is a constant, the Yukawa coupling must scale like $1/\sin\theta_{Sh}$, resulting in
\be
y_Q\approx2.4\times10^{-11}\sqrt{\frac{\lambda_h}{\lambda_{Sh}}}.
\ee
This relation is visible in the small $\lambda_{Sh}$ region of the upper right plot in figure \ref{scatterplots}. Using equations (\ref{eq:pionmass}) to (\ref{eq:vevs}) we can solve for the self coupling of the messenger scalar
\be
\lambda_S=\left(1+\lambda_h\frac{v_{EW}^4}{f_\pi^2m_\pi^2}\right)\frac{\lambda_{Sh}^2}{\lambda_h^2}\approx 108\lambda_{Sh}^2,
\label{eq:lambda_S}
\ee
so that for the region $\lambda_{Sh}<10^{-3}$ the approximation $\lambda_S\ll\lambda_{Sh}$ is well justified.

From the above equations we can solve for the dark technicolor confinement scale
\bea
\Lambda_{\rm TC}&\approx&\left(\frac{1-\lambda_h}{\lambda_h}+\frac{v_{EW}^4}{f_\pi^2m_\pi^2}\right)^{\frac{1}{12}}\frac{10^4\sqrt{f_\pi m_\pi}}{2.9\lambda_h^{\frac14}}\lambda_{Sh}^{\frac13} \nonumber \\
&\approx&1.1\times10^6{\rm GeV}\lambda_{Sh}^{\frac13}.
\eea
Looking at the lower left plot in figure \ref{scatterplots}, we can clearly see this scaling in the small $\lambda_{Sh}$ region. The mass of the messenger scalar can similarly be approximated as
\bea
m_S&=&\frac{\sqrt{\left(\sqrt{3-\lambda_h}{\lambda_h}+\frac{3v_{EW}^4}{f_\pi^2m_\pi^2}\right)f_\pi m_\pi}}{\left(1+\lambda_h\left(\frac{v_{EW}^4}{f_\pi^2m_\pi^2}-1\right)\right)^{\frac14}}\sqrt{\lambda_{Sh}}\nonumber \\
&\approx& 870{\rm GeV}\sqrt{\lambda_{Sh}},
\eea
which in the phenomenologically favoured range of $\lambda_{Sh}$ implies that $m_S$ gets values from \mbox{$\sim10^{-2}$ GeV} to a few hundred GeV, excluding the region near the Higgs mass where the mixing effects are large, as discussed above. Going further towards the high $m_S$ would imply larger values for the scalar couplings $\lambda_{Sh},\lambda_S$, and from equation (\ref{eq:lambda_S}) we can see that $\lambda_S$ soon becomes nonperturbative. Therefore we will not explore this region of the parameter space.

The last plot on the lower right in figure \ref{scatterplots} shows the lifetime of the unstable pion, resulting from the decay width into $b\bar{b}$. Looking at the formula shown in equation (\ref{eq:pion width}) we can see that, since $\sin\theta_{Sh}y_Q$ is a constant, the resulting lifetime is essentially a constant in the small $\lambda_{Sh}$ and hence small $m_S$ region. As $\lambda_{Sh}$ and consequently $m_S$ increases and approaches the value \mbox{$m_S=m_\pi=80$ GeV} there is a resonance in the decay width, resulting in the dip around \mbox{$\lambda_{Sh}\sim0.005$}. As $m_S$ exceeds the Higgs mass, the scattering processes relevant for the freeze-in become dominated by $s$-channel Higgs exchange instead of the messenger scalar, resulting in a slight shift in the required value of $\sin\theta_{Sh}y_Q$ and hence in the pion lifetime seen in the plot. In all the plots shown in figure \ref{scatterplots} the width of the band of accepted results corresponds to the width of the accepted values for the relic density, which we have taken as $(0.5-2)$ times the observed central value, and the width of the accepted values for $f_\pi$, resulting from equation (\ref{eq:self scattering xsec}).

The mass of the messenger scalar is predicted to be quite small, with the preferred parameter space containing values in the range \mbox{$m_S\in [10^{-2},10^3]$ GeV.} However, since the couplings between the messenger scalar and the SM are very weak, this scalar resonance is unobservable at LEP or at the LHC.\\

\section{Discussion and Conclusions} 
\label{sec:conclusions}

In this paper we have proposed a mechanism to generate the galactic centre GeV gamma ray signal from self-interacting composite dark matter in a classically scale-invariant model. Because of the small Higgs portal coupling between the dark and visible sectors, thermal freeze-out is not a viable way to generate the observed relic density, if there is no fine-tuned resonance enhancement of the portal interaction. Instead, we propose that the couplings of the messenger scalar to both sectors are small, such that dark matter can be produced through freeze-in.

We notice that the phenomenologically favoured parameter space contains very small values for all of the couplings of the messenger scalar, $\lambda_S,\lambda_{Sh}, y_Q$, and a moderate hierarchy between the mass scales $\Lambda_{\rm TC}, v_{\rm EW}$. It should be noted, however, that taking all these couplings to zero corresponds to a fixed point in the renormalization group running of the messenger scalar, and results in an enhanced symmetry of the model, since the dark sector becomes completely decoupled from the SM in that limit. As was discussed in \cite{Foot:2013hna}, this scenario is technically natural and does not contain a hierarchy problem.

Due to the composite nature of the dark matter particles, they naturally have sizable self interactions. However, in our model the self-interaction cross section is within $\mathcal{O}(1)$ of the annihilation
cross section, which is too weak to  alleviate the known problems in small scale structure formation. The prospects for direct detection of this type of dark matter particle are weak, since the dark matter-SM scattering cross section is suppressed by the small Higgs portal coupling, and is therefore orders of magnitude below the current experimental sensitivity. Precision studies of the couplings and decay branching ratios of the Higgs boson could perhaps reveal the presence of the messenger scalar in a future electron-positron-collider.

\mysection{Acknowledgments}
This work was supported by  grants MTT60, MJD435, IUT23-6, and by the EU through the ERDF CoE program.


\begin{thebibliography}{00}


\bibitem{Goodenough:2009gk}
  L.~Goodenough and D.~Hooper,
  arXiv:0910.2998 [hep-ph].

\bibitem{Hooper:2010mq}
  D.~Hooper and L.~Goodenough,
  Phys.\ Lett.\ B {\bf 697} (2011) 412
  [arXiv:1010.2752 [hep-ph]].

\bibitem{Boyarsky:2010dr}
  A.~Boyarsky, D.~Malyshev and O.~Ruchayskiy,
  Phys.\ Lett.\ B {\bf 705} (2011) 165
  [arXiv:1012.5839 [hep-ph]].

\bibitem{Hooper:2011ti}
  D.~Hooper and T.~Linden,
  Phys.\ Rev.\ D {\bf 84} (2011) 123005
  [arXiv:1110.0006 [astro-ph.HE]].

\bibitem{Abazajian:2012pn}
  K.~N.~Abazajian and M.~Kaplinghat,
  Phys.\ Rev.\ D {\bf 86} (2012) 083511
  [arXiv:1207.6047 [astro-ph.HE]].

\bibitem{Gordon:2013vta}
  C.~Gordon and O.~Macias,
  Phys.\ Rev.\ D {\bf 88} (2013) 083521
  [arXiv:1306.5725 [astro-ph.HE]].

\bibitem{Abazajian:2014fta}
  K.~N.~Abazajian, N.~Canac, S.~Horiuchi and M.~Kaplinghat,
  Phys.\ Rev.\ D {\bf 90} (2014) 023526
  [arXiv:1402.4090 [astro-ph.HE]].


\bibitem{Hooper:2013rwa}
  D.~Hooper and T.~R.~Slatyer,
  Phys.\ Dark Univ.\  {\bf 2} (2013) 118
  [arXiv:1302.6589 [astro-ph.HE]].

\bibitem{Huang:2013pda}
  W.~C.~Huang, A.~Urbano and W.~Xue,
  arXiv:1307.6862 [hep-ph].


\bibitem{Adriani:2008zq}
  O.~Adriani, G.~C.~Barbarino, G.~A.~Bazilevskaya, R.~Bellotti, M.~Boezio, E.~A.~Bogomolov, L.~Bonechi and M.~Bongi {\it et al.},
  Phys.\ Rev.\ Lett.\  {\bf 102} (2009) 051101
  [arXiv:0810.4994 [astro-ph]].

\bibitem{Adriani:2010rc}
  O.~Adriani {\it et al.}  [PAMELA Collaboration],
  Phys.\ Rev.\ Lett.\  {\bf 105} (2010) 121101
  [arXiv:1007.0821 [astro-ph.HE]].

\bibitem{Adriani:2012paa}
  O.~Adriani, G.~A.~Bazilevskaya, G.~C.~Barbarino, R.~Bellotti, M.~Boezio, E.~A.~Bogomolov, V.~Bonvicini and M.~Bongi {\it et al.},
  JETP Lett.\  {\bf 96} (2013) 621
   [Pisma Zh.\ Eksp.\ Teor.\ Fiz.\  {\bf 96} (2012) 693].


\bibitem{Evoli:2011id}
  C.~Evoli, I.~Cholis, D.~Grasso, L.~Maccione and P.~Ullio,
  Phys.\ Rev.\ D {\bf 85} (2012) 123511
  [arXiv:1108.0664 [astro-ph.HE]].

\bibitem{DiBernardo:2009ku}
  G.~Di Bernardo, C.~Evoli, D.~Gaggero, D.~Grasso and L.~Maccione,
  Astropart.\ Phys.\  {\bf 34} (2010) 274
  [arXiv:0909.4548 [astro-ph.HE]].

\bibitem{Jin:2014ica}
  H.~B.~Jin, Y.~L.~Wu and Y.~F.~Zhou,
  arXiv:1410.0171 [hep-ph].

\bibitem{Daylan:2014rsa}
  T.~Daylan, D.~P.~Finkbeiner, D.~Hooper, T.~Linden, S.~K.~N.~Portillo, N.~L.~Rodd and T.~R.~Slatyer,
  arXiv:1402.6703 [astro-ph.HE].

\bibitem{Hooper:2014ysa}
  D.~Hooper, T.~Linden and P.~Mertsch,
  arXiv:1410.1527 [astro-ph.HE].

\bibitem{Calore:2014xka} 
  F.~Calore, I.~Cholis and C.~Weniger,
  arXiv:1409.0042 [astro-ph.CO].

\bibitem{Basak:2014sza} 
  T.~Basak and T.~Mondal,
  arXiv:1405.4877 [hep-ph].

\bibitem{Modak:2013jya} 
  K.~P.~Modak, D.~Majumdar and S.~Rakshit,
  arXiv:1312.7488 [hep-ph].

\bibitem{Alves:2014yha} 
  A.~Alves, S.~Profumo, F.~S.~Queiroz and W.~Shepherd,
  arXiv:1403.5027 [hep-ph].

\bibitem{Alvares:2012qv} 
  J.~D.~Ruiz-Alvarez, C.~A.~de S.Pires, F.~S.~Queiroz, D.~Restrepo and P.~S.~Rodrigues da Silva,
  Phys.\ Rev.\ D {\bf 86}, 075011 (2012)
  [arXiv:1206.5779 [hep-ph]].

\bibitem{Hur:2007uz} 
  T.~Hur, D.~-W.~Jung, P.~Ko and J.~Y.~Lee,
  Phys.\ Lett.\ B {\bf 696}, 262 (2011)
  [arXiv:0709.1218 [hep-ph]].

\bibitem{Hur:2011sv} 
  T.~Hur and P.~Ko,
  Phys.\ Rev.\ Lett.\  {\bf 106}, 141802 (2011)
  [arXiv:1103.2571 [hep-ph]].

\bibitem{Heikinheimo:2013fta} 
  M.~Heikinheimo, A.~Racioppi, M.~Raidal, C.~Spethmann and K.~Tuominen,
  arXiv:1304.7006 [hep-ph].

\bibitem{Holthausen:2013ota} 
  M.~Holthausen, J.~Kubo, K.~S.~Lim and M.~Lindner,
  JHEP {\bf 1312}, 076 (2013)
  [arXiv:1310.4423 [hep-ph]].

\bibitem{GellMann:1968rz} 
  M.~Gell-Mann, R.~J.~Oakes and B.~Renner,
  Phys.\ Rev.\  {\bf 175}, 2195 (1968).


\bibitem{Buckley:2012ky}
  M.~R.~Buckley and E.~T.~Neil,
  Phys.\ Rev.\ D {\bf 87} (2013) 4,  043510
  [arXiv:1209.6054 [hep-ph]].

\bibitem{Abdullah:2014lla} 
  M.~Abdullah, A.~DiFranzo, A.~Rajaraman, T.~M.~P.~Tait, P.~Tanedo and A.~M.~Wijangco,
  Phys.\ Rev.\ D {\bf 90}, 035004 (2014)
  [arXiv:1404.6528 [hep-ph]].

\bibitem{Berlin:2014pya}
  A.~Berlin, P.~Gratia, D.~Hooper and S.~D.~McDermott,
  Phys.\ Rev.\ D {\bf 90} (2014) 015032
  [arXiv:1405.5204 [hep-ph]].

\bibitem{Hooper:2012cw}
  D.~Hooper, N.~Weiner and W.~Xue,
  Phys.\ Rev.\ D {\bf 86} (2012) 056009
  [arXiv:1206.2929 [hep-ph]].

\bibitem{Ko:2014gha}
  P.~Ko, W.~I.~Park and Y.~Tang,
  JCAP {\bf 1409} (2014) 013
  [arXiv:1404.5257 [hep-ph]].

\bibitem{Boehm:2014bia}
  C.~Boehm, M.~J.~Dolan and C.~McCabe,
  Phys.\ Rev.\ D {\bf 90} (2014) 023531
  [arXiv:1404.4977 [hep-ph]].

\bibitem{Martin:2014sxa}
  A.~Martin, J.~Shelton and J.~Unwin,
  arXiv:1405.0272 [hep-ph].

\bibitem{Cline:2014dwa}
  J.~M.~Cline, G.~Dupuis, Z.~Liu and W.~Xue,
  JHEP {\bf 1408} (2014) 131
  [arXiv:1405.7691 [hep-ph]].

\bibitem{Ko:2014loa}
  P.~Ko and Y.~Tang,
  arXiv:1407.5492 [hep-ph].

\bibitem{Freytsis:2014sua}
  M.~Freytsis, D.~J.~Robinson and Y.~Tsai,
  arXiv:1410.3818 [hep-ph].

\bibitem{Barate:2003sz} 
  R.~Barate {\it et al.}  [LEP Working Group for Higgs boson searches and ALEPH and DELPHI and L3 and OPAL Collaborations],
  Phys.\ Lett.\ B {\bf 565}, 61 (2003)
  [hep-ex/0306033].

\bibitem{Abbiendi:2002qp} 
  G.~Abbiendi {\it et al.}  [OPAL Collaboration],
  Eur.\ Phys.\ J.\ C {\bf 27}, 311 (2003)
  [hep-ex/0206022].

\bibitem{Hochberg:2014dra}
  Y.~Hochberg, E.~Kuflik, T.~Volansky and J.~G.~Wacker,
  arXiv:1402.5143 [hep-ph].

\bibitem{Chu:2011be} 
  X.~Chu, T.~Hambye and M.~H.~G.~Tytgat,
  JCAP {\bf 1205}, 034 (2012)
  [arXiv:1112.0493 [hep-ph]].

\bibitem{Heikinheimo:2013cua} 
  M.~Heikinheimo, A.~Racioppi, M.~Raidal and C.~Spethmann,
  Phys.\ Lett.\ B {\bf 726}, 781 (2013)
  [arXiv:1307.7146].

\bibitem{Foot:2013hna} 
  R.~Foot, A.~Kobakhidze, K.~L.~McDonald and R.~R.~Volkas,
  Phys.\ Rev.\ D {\bf 89}, 115018 (2014)
  [arXiv:1310.0223 [hep-ph]].





\end{thebibliography}
\end{document}